\documentclass[final,conference]{IEEEtran}
\usepackage{cite}
\usepackage{amsmath,amssymb,amsfonts}
\usepackage{graphicx}
\usepackage{textcomp}
\usepackage{xcolor}
\def\BibTeX{{\rm B\kern-.05em{\sc i\kern-.025em b}\kern-.08em
    T\kern-.1667em\lower.7ex\hbox{E}\kern-.125emX}}

\usepackage{pgfplots}
\usepackage{pgfplotstable}
\usepackage{filecontents}
\usepgfplotslibrary{groupplots}
\pgfplotsset{compat=1.17} 

\usepackage{booktabs}
\usepackage{amsfonts}
\usepackage{amsmath}
\usepackage{bbm}
\usepackage{hyperref}
\usepackage[capitalize]{cleveref}
\usepackage[normalem]{ulem}
\usepackage{soul}
\usepackage{xcolor}
\usepackage[inline]{enumitem}
\usepackage{siunitx}
\usepackage{float}

\usepackage[textsize=tiny, textwidth=1.3cm]{todonotes}
\usepackage{subcaption} 
\usepackage{color, colortbl}
\usepackage{bbold}
\usepackage{tikz}
\usepackage{algorithm}
\usepackage{url}
\usepackage{algpseudocode}
\usepackage{mathtools}
\usepackage{makecell}
\usepackage{graphicx}
\usepackage{flushend}
\useunder{\uline}{\ul}{}

\usepackage{tabularx}

\DeclareMathOperator*{\argmax}{argmax}

\newcommand*{\argmaxl}{\argmax\limits}
\algnewcommand\algorithmicforeach{\textbf{for each}}
\algdef{S}[FOR]{ForEach}[1]{\algorithmicforeach\ #1\ \algorithmicdo}

\renewcommand{\Cref}[1]{\cref{#1}}
\Crefname{equation}{Eq.}{Eqs.}
\Crefname{figure}{Fig.}{Figs.}
\Crefname{tabular}{Tab.}{Tabs.}
\Crefname{table}{Tab.}{Tabs.}
\Crefname{algorithm}{Algo.}{Algos.}

\begin{document}

\title{An Online Approach to Solving Public Transit Stationing and Dispatch Problem}

\makeatletter
\newcommand{\linebreakand}{%
  \end{@IEEEauthorhalign}
  \hfill\mbox{}\par
  \mbox{}\hfill\begin{@IEEEauthorhalign}
}
\makeatother

\author{\IEEEauthorblockN{Jose Paolo Talusan}
\IEEEauthorblockA{\textit{Vanderbilt University} \\
Nashville, USA \\
jose.paolo.talusan@vanderbilt.edu}
\and
\IEEEauthorblockN{Chaeeun Han}
\IEEEauthorblockA{\textit{Pennsylvania State University} \\
University Park, PA, USA \\
cfh5554@psu.edu}
\and
\IEEEauthorblockN{Ayan Mukhopadhyay}
\IEEEauthorblockA{\textit{Vanderbilt University} \\
Nashville, USA \\
ayan.mukhopadhyay@vanderbilt.edu}
\linebreakand
\IEEEauthorblockN{Aron Laszka}
\IEEEauthorblockA{\textit{Pennsylvania State University} \\
University Park, PA, USA \\
alaszka@psu.edu}
\and
\IEEEauthorblockN{Dan Freudberg}
\IEEEauthorblockA{\textit{WeGo Public Transit} \\
Nashville, USA \\
dan.freudberg@nashville.gov}
\and
\IEEEauthorblockN{Abhishek Dubey}
\IEEEauthorblockA{\textit{Vanderbilt University} \\
Nashville, USA \\
abhishek.dubey@vanderbilt.edu}
}


\maketitle

\begin{abstract}
Public bus transit systems provide critical transportation services for large sections of modern communities. On-time performance and maintaining the reliable quality of service is therefore very important. Unfortunately, disruptions caused by overcrowding, vehicular failures, and road accidents often lead to service performance degradation. Though transit agencies keep a limited number of vehicles in reserve and dispatch them to relieve the affected routes during disruptions, the procedure is often ad-hoc and has to rely on human experience and intuition to allocate resources (vehicles) to affected trips under uncertainty. In this paper, we describe a principled approach using non-myopic sequential decision procedures to solve the problem and decide (a) if it is advantageous to anticipate problems and proactively station transit buses near areas with high-likelihood of disruptions and (b) decide if and which vehicle to dispatch to a particular problem. Our approach was developed in partnership with the Metropolitan Transportation Authority for a mid-sized city in the USA and models the system as a semi-Markov decision problem (solved as a Monte-Carlo tree search procedure) and shows that it is possible to obtain an answer to these two coupled decision problems in a way that maximizes the overall reward (number of people served). We sample many possible futures from generative models, each is assigned to a tree and processed using root parallelization. We validate our approach using 3 years of data from our partner agency. Our experiments show that the proposed framework serves 2\% more passengers while reducing deadhead miles by 40\%.
\end{abstract}


\section{Introduction}

There is a growing demand from people living in urban and metropolitan areas for better and more efficient public transit systems~\cite{status_highways_2003}. Public transit ridership has risen to more than 70\% of pre-pandemic levels, signifying an upward trend~\cite{ridership_report_2022}. Thus, public transit agencies are now encountering more incidents of crowding, vehicular failures, and road accidents which lead to service performance degradation. 
In 2022, the partner agency had approximately 100 buses in their fleet, each one assigned an average of $15$ trips, to serve hundreds of passengers every day. With such high usage rates, accidents and mechanical failures are bound to happen. In 2022 alone, there were over 6500 reports of service disruptions due to a variety of factors such as weather, accidents, or mechanical issues. These issues, along with passenger overcrowding and delays, are some of the things that can degrade the transit's quality of service, eroding trust and eventually leading to a decrease in usage.

In response to these issues, transit agencies often have a limited pool of vehicles used to cover for and mitigate the potential loss of service that these disruptions would cause. Thus, the task of optimal allocation and dispatch of these limited resources, a fundamental problem faced by various agencies, lies upon the operators and transit experts. While their decisions may reflect the best solutions at that time, in the end, their choices are myopic and based on domain knowledge.

We identify this as a \textbf{dynamic scheduling and dispatch for fixed-line transit problem}. Transportation agencies traditionally are expected to have a fixed schedule with regular intervals, meant to provide wide-area coverage. However, the need for dynamic scheduling arises when dealing with origin-destination passenger demands or paratransit services, where demands along a route vary significantly. In such cases, fixed routes and schedules cannot be used to efficiently model the system. The results are flexible routing and scheduling strategies such as deadheading or dynamic stop skipping, to determine which vehicle to send and which stops to skip~\cite{fu_real-time_2003}. 
Solving this problem in the real world, where future events are considered, requires solving hard combinatorial optimization problems. This is a computationally intractable problem, considering the number of buses, stops, and events that one must take into account.
In light of these challenges, the implementation of dynamic scheduling and dispatch solutions has gained considerable attention within the public transit domain. These systems leverage advanced data analytics and predictive modeling techniques to address the complex scheduling issues faced by transit agencies. By integrating real-time data streams and historical performance metrics, these solutions enable transit authorities to make informed decisions to enhance the efficiency and reliability of their services.


With the integration of cutting-edge technologies and data-driven insights, transit agencies are better equipped to optimize their resources and respond promptly to disruptions. By embracing dynamic scheduling and dispatch solutions, these agencies can provide more agile and adaptable public transit services, ultimately improving urban transportation systems' overall quality and reliability.

%

\textbf{Contributions}: Our approach to solving this problem is to model it as a sequential decision-making problem under uncertainty~\cite{Kochenderfer_2015}, which can be solved to maximize some domain-specific utility function such as the total number of passengers served~\cite{jung_dynamic_2016,newell_1971}. Our main contributions are summarized as follows: \textbf{(1)} we design a fully online and non-myopic solver for stationing and dispatch for fixed-line transit services, which aims to optimize limited resources in the form of substitute buses to maximize the number of passengers served. \textbf{(2)} We model this problem as a semi-Markov decision process which we solve using MCTS. \textbf{(3)} Extensive experiments are conducted on both generated and real-world public transit datasets, and the results show that our model outperforms a greedy baseline model. Increasing total passengers served by as much as 7\% and reducing deadhead miles by 42\%. This shows that there is room for improvement in the current methods used by certain public transit agencies.


\section{Problem Statement}
\label{sec:problem_statement}

\begin{figure*}[t!]
    \centering
    \begin{subfigure}[t]{0.55\textwidth}
        \centering
        \includegraphics[height=1.6in,width=3.8in,keepaspectratio]{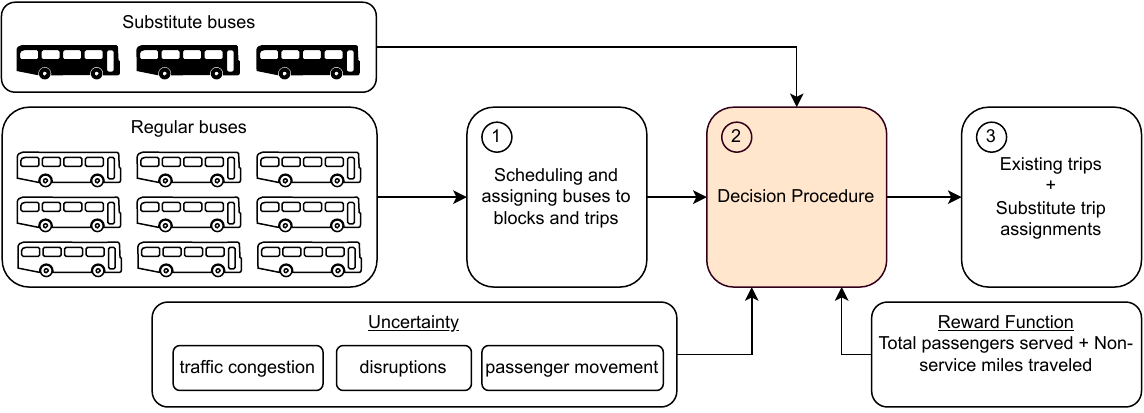}
        \caption{Overview}
        \label{fig:overview}
    \end{subfigure}%
    ~ 
    \begin{subfigure}[t]{0.35\textwidth}
        \centering
        \includegraphics[height=1.7in]{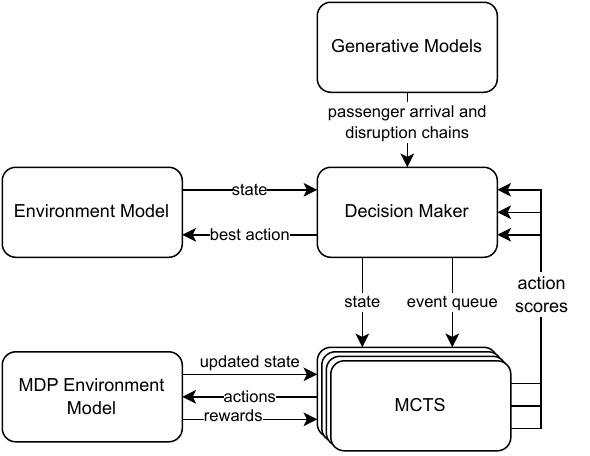} 
        \caption{Decision Procedure}
        \label{fig:framework}
    \end{subfigure}
    \caption{(a) An overview of the stationing and dispatch problem. \textcircled{1}Transit agencies have a set schedule for their buses as detailed in GTFS. A subset of buses is held back, to be dispatched in case of incidents, based on some \textcircled{2} decision procedure and some desired reward. Finally, \textcircled{3} a list of assignments for all buses is generated. (b) Our proposed approach runs inside \textcircled{2} Decision Procedure. It creates generative models based on real-world data and outputs the optimal actions for trip assignments.}
\end{figure*}

In this section, we describe the characteristics of fixed-line transit and formulate the dynamic stationing and dispatch problem. We then discuss the concepts of stationing and dispatch, introduce the action spaces and finally model the problem as a semi-Markov decision process. We present a notation lookup table in the technical appendix (\cref{tab:symbols}) for convenient reference to notation.

\subsection{Fixed-line Transit and Substitute Buses}
Fixed-line public transit relies on a set of vehicles following a consistent and predictable schedule of trips along a set of routes. Any events or incidents that would cause buses to deviate from their schedule can lead to increased passenger wait times, headway bunching, interfere with transfers, and eventually impact reliability and erode passenger trust. In response, the agency has a limited number of substitute buses, without any scheduled trips, that can be deployed to minimize the effect of unforeseen events.

\textbf{Trip:} is a sequence of locations or stops that a bus will visit along its route. Each bus starts and ends its service day at a \textbf{depot}, which are central system hub. Formally, a trip $i$ is defined by an ordered sequence of stops $T^i = \{L^i_0, L^i_1, L^i_2, \dots L^i_n\}$ where $L^i_0$ and $L^i_n$ are the depots if the vehicle is traversing a regular planned trip or staging areas if the vehicle was dispatched. Each stop has a predetermined scheduled arrival time along the trip. 

\textbf{Route + Direction}: A bus route is a group of trips that a bus is expected to follow recurrently. A regular bus assigned to a particular route will travel in a certain direction until it turns around for the next trip going back. Regular buses will not diverge from assigned routes.

\textbf{Stops:} $j$ in a trip $i$ have a predetermined scheduled time when a vehicle is expected to arrive, denoted by $e(L^i_j)$. However, buses can encounter incidents along their trip that may cause delays or disruptions, preventing passengers from boarding the bus on time or preventing them from boarding at all, leading to dissatisfaction and frustration. If a bus arrives at a stop at or near full capacity, then some passengers will not be able to board leading to an \textbf{overage event}. On the other hand, if a bus breaks down or encounters an incident en route it leads to an \textbf{disruption event}.

For a regular service day, the transit agency has access to two types of buses, regular and substitute. \textbf{Regular buses} travel along their assigned routes for the entire service day. Regular buses do not diverge from their schedules and they cannot be used for other trips even when they are currently idle (between trips). In preparation for incidents, agencies may have several \textbf{substitute buses} that are kept apart from regular fixed-line services and are sent out in response to potential service disruptions such as overage events, delays, and vehicle breakdowns. Substitute buses can be \textit{stationed} to certain stops along the city, effectively waiting until it has to be \textit{dispatched} to either takeover a broken down regular bus or pick up passengers left behind by one. The main objective for using substitute buses is to maximize the number of passengers served and minimize passenger wait time at the bus stops. Thus, it is important to station vehicles at locations close to routes and stops that frequently encounter issues, thereby maximizing substitute bus efficiency and decreasing response times. The efficiency of a substitute bus can be measured by the number of passengers it has served and the amount of \textit{deadhead miles} it has traveled. Deadhead miles constitute the distance a substitute bus has to travel between where it is stationed to the dispatch location, during which they are not accepting any passengers. If there are no passenger overages or disruption incidents, then idle substitute buses can be re-stationed in preparation for future events. Ideally, there would always be a vehicle that can be dispatched to cover any affected trip at any time. However, resource constraints in both manpower and budget limit the total number of buses that can be used as substitute buses.
Therefore, a decision must be made. There are two types of actions the decision-maker, in this transit supervisors, can optimize: (1) which substitute buses to send to cover for trips (\textit{dispatching actions}) and (2) which stops or depots to stage an idle bus in anticipation of future events (\textit{stationing actions}). This dynamic stationing and dispatch problem is visualized in \cref{fig:overview}, which is an extension of the traditional, fixed-line transit.

Formally, the dynamic stationing and dispatch fixed-transit problem consists of a set of vehicles denoted as $V$. A subset of vehicles are used to service the regular fixed-line transit schedule for the day, denoted as $V^R \subset V$ and the remaining vehicles are substitute vehicles available for stationing and dispatching $V^O = V \setminus V^R$. substitute buses start stationed at predefined stationing areas located throughout the geographic region of operation.  
While the state space in this stationing and dispatch problem is in continuous time, it is convenient to view the system's dynamics as a set of finite decision-making states that evolve in discrete time. For example, a bus moving between stops on its trip continuously changes the state of the world but presents no scope for decision-making unless an event occurs that needs a response (such as a bus breaking down or passengers getting left behind) or the planner decides to re-station the buses. As a result, the decision-maker only needs to find optimal actions for a subset of the state space that provides the opportunity to take actions. 

We define this moment of decision-making as a decision epoch. A decision epoch is triggered each time a bus arrives at a stop, has an overage or disruption event, or if a stationing event is triggered. At each decision epoch, there are three possible actions for each idle substitute vehicle: (1) They can be stationed and moved to \textbf{any} stop or depot across the region, called deadheading~\cite{fu_real-time_2003}, (2) They can be dispatched to a stop on any trip. They must serve this entire trip, serving all subsequent stops, before becoming available for dispatch again, or (3) They can be made to wait at their current location.

\subsection{Fixed-line Stationing \& Dispatch as SMDP}

To solve the problem of identifying the best assignments for substitute buses, we model the dynamic stationing and dispatching problem as a semi-Markov decision process (SMDP). We treat the model as semi-Markovian since buses must physically move from stop to stop along their trip or from staging areas when dispatched or reallocated. Travel time during these transitions are inconsistent due to latent factors such as traffic congestion, passenger behavior, or driver actions, which cause the temporal transitions between states to be non-memoryless. 

An SMDP can be represented as the tuple $\{S,A,P,T, R, \gamma\}$ where $S$ is a finite state space,  $A$ is the set of actions, $P$ is a state transition function, $T$ is the temporal distribution over transitions between states, $R$ is an instantaneous reward function, and $\gamma$ is the discount factor for future rewards~\cite{hu2007markov}.

\textbf{States:} 
We denote the set of states as $S$. 
A state at time $t$, denoted by $s_t$, consists of a tuple $(V_t, T_t, vloc_t, V^s_t)$ where $V_t$ is the set of all vehicles, $T_t$ is the set of all trips, $vloc_t$ is the location of all vehicles, and $V^s_t$ is the state of each vehicle. 
We associate some additional information with each vehicle: $c(V^i)$ is the capacity of vehicle $i$, $o(V^i_t)$ is the current occupancy of vehicle $i$ at $t$, and $w(V^i_t)$ is the status of the vehicle.
The status of a vehicle can be \textit{in transit} if it is actively servicing a route, \textit{idle} if it is sitting at a staging location ready to be dispatched, or \textit{out-of-service} if it is no longer available. We assume that no two events occur simultaneously in our system model. If such a case arises, we can add an arbitrarily small time interval to separate the two events, creating separate states.

\textbf{Actions:} We denote the set of all feasible actions at time $t$ by $A^t$.
There are three types of possible actions for substitute buses: stationing, dispatch, and no action. 

\textit{Dispatch actions} selects an idle substitute vehicle and assigns it a trip. Thus, requiring them to be dispatched from their current location to a target stop along the assigned trip. A dispatch action ($A_d$) selects a vehicle $i$, trip $j$, and stop $k$ along that trip given the current state of the system and can be denoted as $A_d: S \mapsto \{V^i, T^j, L^k\}$ where $V^i \in V^O$ and $L^k \in T^j$. The substitute bus must now visit all stops from $L^k$ to $L^n$, where $n$ is the last stop for trip $j$. The goal of dispatch actions is to directly alleviate excess demand from the system, by replacing broken vehicles or increasing capacity for crowded trips.

\textit{Stationing actions} reallocates idle substitute buses by moving them between different stops around the city. The goal is to preemptively place idle vehicles closer to trips and stops that may encounter issues in the future. Stationing rebalances idle vehicles in anticipation of expected future demand. Therefore, a reallocation action can be formulated as $A_r: S \mapsto \{V^i, D^j\}$ where $V^i \in V^O$ and $D^j$ is one of the staging locations.

Otherwise, the system can simply \textit{Do nothing}. In this case, no dispatching or reallocation actions are chosen and the system continues as is. 


\textbf{State Transitions:} 
At each decision epoch $t$, the decision maker can take an action such as $a^j \in A_t$, which will transition from the pre-decision state $s_t$ to a post-decision state $s'_t$. 
Afterward, the post-decision state $s'_t$ transitions into the next pre-decision state $s_{t+1}$, within which new events will occur. 
During state transitions, buses travel between stops along their assigned trips, drop off passengers waiting to alight, and pick up any passengers waiting at the stop. 
Passengers waiting at a bus stop will get on the bus that arrives, provided that it is going along their desired route and direction, and if the bus is not yet over capacity $o(V^i_t) < c(V^i)$
Passengers who are not able to get on the bus will wait for another bus for a certain amount of time before leaving. Buses in transit can also encounter a disruption event while traveling between stops. We refrain from discussing the mathematical model and expressions for the temporal transitions and the state transition probabilities $\mathcal{P}$ since our algorithmic framework relies only on a generative model of the world and not explicit estimates of the transitions themselves.

We model the stop-to-stop travel times of buses by sampling an independent empirical distribution based on historical data. The number of boarding and alighting passengers at every stop is generated by generative models which will be discussed in the next section. Finally, disruptions are based on a statistical model trained to forecast incidents.

\textbf{Rewards:} The reward for an action $a^j \in A_t$, defined as $\gamma(a^j)$, is based on the total number of people that successfully get on the bus and the non-service miles traveled by substitute buses.
It only considers any actions that maximize the number of people that a limited number of substitute buses can pick up while minimizing the non-service miles traveled.
%
The decision maker's goal is to find an optimal policy that maps system states to actions to maximize long-term utility.
In our case, the utility is related to reducing the overcrowding of the fixed-line transit system and mitigating the effects of vehicle breakdowns, traffic incidents, and passenger overages.

\section{Approach}
\label{sec:approach}
The dynamic environment present in our problem discourages the use of typical offline-online solutions. Before being embedded in an online search, offline components require long training times and must be re-trained each time the environment changes. This motivates us to use Monte-Carlo Tree Search (MCTS). MCTS is a simulation-based search algorithm that has been widely used in game-playing scenarios. MCTS-based algorithms evaluate actions by sampling from a large number of possible scenarios. Being an anytime algorithm, MCTS can immediately incorporate any changes in the underlying environment when making decisions.
We show the framework for our proposed approach in \cref{fig:framework}. 
All of our code for the implementation is available at \url{https://zenodo.org/records/10594255}.

\begin{table}[!t]
\scriptsize
\centering
    \caption{Detail of datasets}
    \label{tab:dataset_details}
    \begin{tabular}{|c|c|c|c|c|}
    \hline
    \textbf{Data} & \textbf{Size} & \textbf{Rows} & \textbf{Features} & \textbf{Description} \\   
    \hline
    \makecell{Automatic\\Passenger\\Count (APC)} & 
        3.3GB &
        17M &
        \begin{tabular}{@{}c@{}} Date \\ Load \\ Arrival time \\ Departure time \\ \end{tabular} &
        \begin{tabular}{@{}c@{}} Date of trip \\ Occupancy at stop \\ Arrival at stop \\ Departure from stop \end{tabular}\\
    \hline
    Incidents & 
        344MB &
        2.3M &
        \begin{tabular}{@{}c@{}} Date \\ Type \\ Confidence \\ Location \\ \end{tabular} &
        \begin{tabular}{@{}c@{}} Date of trip \\ Occupancy at stop \\ Number of reports \\ Coordinates \end{tabular}\\
    \hline
    Disruptions & 
        800KB &
        6534 &
        \begin{tabular}{@{}c@{}} Date \\ Type \\ Last stop \\ Location \\ \end{tabular} &
        \begin{tabular}{@{}c@{}} Date of trip \\ Agency classification \\ Last visited stop \\ Coordinates \\ \end{tabular}\\
    \hline
    Road Network & 
        31MB &
        7414 &
        \begin{tabular}{@{}c@{}} Network graph \\ Geometry \\ \end{tabular} &
        \begin{tabular}{@{}c@{}} OSM network \\ Road Segment \end{tabular}\\
    \hline
    \multicolumn{5}{p{6cm}}{Note: All data are from 01-01-2020 to 12-31-2022.}
    \end{tabular}
\end{table}

\subsection{Available Data}

Our solution for this problem is largely data-driven and we have been continuously collecting various datasets shown in \cref{tab:dataset_details}. Automatic Passenger Counts (APC) are data gathered by electronic devices installed on public transit buses. They log the number of passengers boarding and alighting through the use of door sensors. They also collect the position of buses, as well as their arrivals and departures from stops, allowing stop-to-stop travel times to be derived. APC data is not as precise as data gathered by smart cards~\cite{MA20131}. Thus, they require cleaning and augmentation before they can be used for training forecasting models~\cite{apc_data}. 


We also collect both traffic incidents and bus failures, gathered by Waze and the transit agency respectively, to further improve our simulation of the real-world environment. However, these data do not neatly fit with APC data. We have to create a buffer in both space and time when joining them with bus APC data. Thus, they are simply approximations of the real world. Finally, the need to collect all of this data is further necessitated by the fact that APC does not collect origin-destination (OD) data. Due to the way they collect data, they do not have specific information regarding which stop a particular passenger got on and at which stop they got off.


\begin{algorithm}[!t]
    \small
    \begin{algorithmic}[1]
        \Require {$e(L), features$}
        \State $E \gets \{\}$
        \ForEach {$(i, j) \in e(L)$}
            \State $p = predict(e(L_j^i, features)$ \Comment{using model from~\cite{apc_data}}
            \State $bin = randomBinFromProbability(p)$
            \State $\hat{e(L_j^i)} = randomInt(bin_{low}, bin_{hi})$
            \State $E(e(L_j^i)) \gets \hat{e(L_j^i)})$
        \EndFor
        \State \Return $E$
    \end{algorithmic}
    \caption{Passenger Distribution Model Generation}
    \label{algo:generative_model_gen}
\end{algorithm}


\subsection{Generative Models}
\label{subsec:generative_model}

Modeling the transition space and representing the stochasticity in our environment necessitates the ability to sample travel times, passenger arrival times, and potential disruptions. 


\textbf{Passenger arrivals:} 
Due to the limitations in the APC data, we are unable to identify passenger counts and waiting times at stops and which stops are their exact destinations. Having only access to current bus occupancy at every stop, we instead computed the boarding and alighting counts by training and sampling a model for forecasting passenger occupancy across stops within a single trip. \cref{algo:generative_model_gen} details how these counts are calculated. For each stop, we forecast probabilities that the occupancy will be in a certain bin. We select a bin based on these probabilities (line 4) and then from within the bin, we select uniformly at random an occupancy level(line 5). We then use this predicted occupancy to generate stop-level boarding and alighting numbers, providing a learned estimate for passenger counts along stops for a given route, based on other features such as weather, time of day, day of the week, and special holidays. Only boardings are allowed at the start of trips and passengers are all required to alight at the stop at the end of every trip. Stochasticity is introduced by sampling multiple chains of events for each service day and by sampling passenger arrival times from a uniform distribution ranging from some time to the time before the bus arrives.


\textbf{Disruption forecasting:}
We forecast the likelihood of disruptions occurring for certain trips by treating it as a supervised learning problem. We train an XGBoost model to predict the binary outcome of the presence of disruptions given a vector of features. Due to the inherent sparsity of available data, the forecast is limited to predicting disruptions at the trip level. However, due to how the problem is modeled, disruptions need to be predicted to the precision of a specific stop at a given time. We generate a simple probability distribution for all the stops along every trip using empirical disruption data. Thus, we predict disruptions using a two-step process, first by identifying whether each trip for the day has a disruption and then identifying the most probable stop where the disruption might occur.

\begin{figure*}[!t]
    \centering
    \includegraphics[height=1.8in]{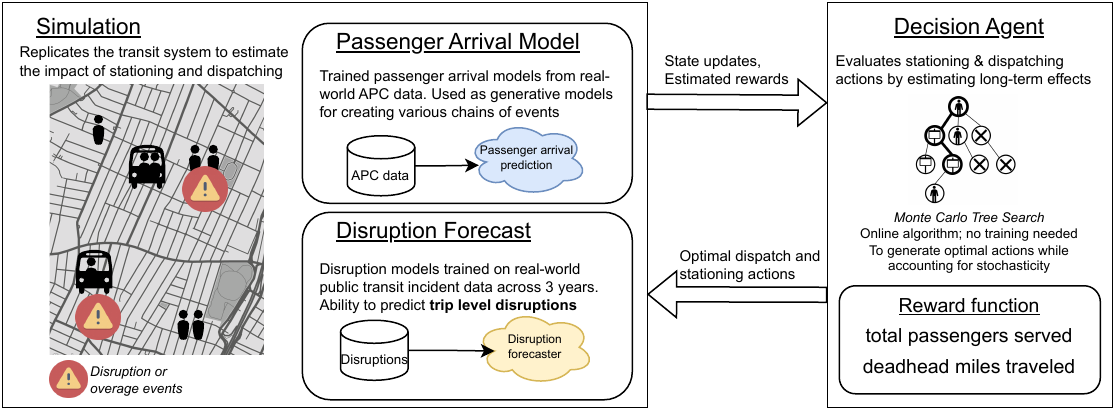}
    \caption{Simulator diagram depicting the main components, passenger arrival model, disruption forecast, and decision agent.}
    \label{fig:sim_diagram}
\end{figure*}

\subsection{Event-Driven Simulator}
The simulator recreates an entire service day using all the prior datasets and models. There are three main components: \textit{buses, passenger arrivals, and disruptions}, shown in~\cref{fig:sim_diagram}. All regular buses are assumed to start the day at the first stop for their first trip. Meanwhile, substitute buses start the day at the main garage. Any movement by substitute buses where they are not assigned any trips adds up to their deadhead miles. Regular buses have no deadhead miles. 

This is an event-driven simulator, events are based on the bus' arrival at stops. Upon arriving at a stop, an active bus assigned to a trip will pick up passengers currently waiting at the stop. Since there are several stops shared by different routes and directions, passengers will only board buses that are headed in the same route and direction that they are going to. Buses only have a limited capacity for passengers. An \textit{overage event} is triggered when passengers cannot board the bus due to overcapacity. They then stay at the bus stop and wait for some determined amount of time before eventually leaving. They will wait for the next bus that will pass by after some headway duration. Passengers can only board at the start of the trip and are required to alight at the end of the trip unless the current trip is the first part of a multi-stage trip where passengers can choose to remain. At any time during the journey, a regular bus may encounter a \textit{disruption event} that causes it to break down and be unable to continue on any of its assigned trips. At this moment, all passengers on the broken bus are assumed to transfer back to the last passed stop. These passengers will wait for the next bus and leave after a set amount of time.

During any overage or disruption events, if there are available substitute buses, the decision-maker can opt to dispatch one to service the stranded passengers. If a substitute bus is sent to cover for a broken-down bus, it travels from its current location to the last passed stop, accruing deadhead miles in the process. Once it reaches the stop, it takes over all the trips assigned to the previous bus and then fetches any passenger waiting at the stop (both stranded and newly arrived ones). This substitute bus is then locked to serve the remaining trips as the original broken bus. The original bus is considered broken for the entirety of the simulation. Once this substitute bus is finished with the assigned trips, it remains at the last stop, waiting to either be stationed or dispatched again. However, if a substitute bus is sent to recover stranded passengers due to overage events, it travels from its current location to the stop where the passengers are. From there. it will only visit all subsequent stops the original bus was assigned to for that given trip that caused the overage event. After reaching the terminal stop of this single trip, it then waits to be stationed or dispatched.

\subsection{Decision Algorithm} 
In the simulator, optimal decisions are taken by an MCTS algorithm. Here stationing and dispatch problems are represented as game trees, where each node in the tree represents a state and each edge is an action that transitions from one state to the next. The current state of the environment is set as the tree's root node and the tree is incrementally expanded and asymmetrically explored. The asymmetric exploration of MCTS allows it to favor more promising nodes while other nodes still have a non-zero probability of being selected, this allows for relatively fast exploration of large action spaces. The value of an action is estimated by simulating a ``rollout'' to the end of a planning horizon while selecting actions based on a default policy. 
%

In the simplest case, the default or rollout policy is computationally cheap such as selecting actions in a uniformly random manner. The tree is grown incrementally while adding and evaluating nodes and their utility until some computational budget has been reached. The algorithm then terminates and returns the best action for the current state. MCTS is a heuristic algorithm~\cite{browne_survey_2012} that requires very little domain-specific knowledge to get acceptable performance. For our stationing and allocation problem, we provide an environmental model, a tree policy for navigating the search tree, and the default policy for producing a value estimate. 

For every decision epoch, an asymmetric search tree is generated by MCTS to obtain the best action for any state. We describe below the flow of our MCTS implementation and the domain-specific components required at each step. The service day starts when buses start their first trip. 
The event chains constructed by the generative model (Sec.~\ref{subsec:generative_model}) provide the number of passengers waiting at the first stops for these initial trips. We use an empirical model to sample the travel times between stops based on historical data. As a bus travels between stops, we sample from a distribution the probability of a trip getting disrupted either due to mechanical issues, traffic accidents, or passenger-related incidents. 

\textbf{Constraints on Decisions} To ensure tractability and to limit the number of decision epochs, due to the large amount of times buses will arrive at stops, we impose a strict interval between decision epochs. 
A bus arriving at a stop will only be considered for decision epoch for dispatch actions if it has been $N$ minutes since the last time the particular bus has been considered as a decision epoch. Independent of this, a decision epoch for stationing actions will occur every $M$ minutes from the start to the end of the service day.


\textbf{Constraints on Actions} Upon reaching a decision epoch, there are only a limited amount of actions that are considered. For stationing decisions, then any idle substitute buses are considered for reallocation to any of the possible stationing locations. For dispatch decisions triggered by an overage event, we consider dispatching to all the stops that have been visited by the current bus on their assigned trip. This includes stops that have stranded passengers as well as the current stop the bus is currently on. We also assign a constraint that trips can only be serviced by one substitute bus. At each decision epoch, only a single substitute bus will be considered for any possible action. For decision epochs where there are no substitute buses available, no action will be taken. Finally, any stops that have been last visited by a recently broken-down bus will be considered in the action space. We use the standard Upper Confidence Bound for Trees (UCT)~\cite{kocsis_bandit_2006} to navigate the search tree and decide which nodes to expand:
\begin{equation}
    \text{UCT} = \hat{X}_j + C_p\sqrt{\frac{\ln n}{n_j}}
\end{equation}
where $n$ is the number of times the current parent node has been visited, $n_j$ is the number of times child $j$ has been visited and $C_p > 0$ is a constant. Ties among child nodes are broken randomly. $\hat{X}_j$ is the estimated utility of state at node $j$.

When working outside the MCTS tree to estimate the value of an action during rollout, we rely on a default policy. This lightweight policy is simulated up to a time horizon, with resulting utility being propagated up the tree. The default policy only considers dispatching and it always selects the nearest idle substitute bus for dispatching.


\textbf{Rewards} is a function of the total served passengers and total deadhead miles for all substitute buses. A trip may be serviced by both a regular and substitute bus. Values are normalized by the total number of passengers (served plus those left behind) and total distance traveled by regular buses. Thus, the reward is calculated as:
\begin{equation}
    \gamma = \alpha\sum_{i\in T}\sum_{j \in L} b(L^i_j) + \beta\sum_{v\in\mathcal{V}^o} d(v)
\end{equation}
where $b(L^i_j)$ is the number of passengers who boarded the bus at stop $j$ on trip $i$ and $d(v)$ is the deadhead miles traveled by bus $v$.

\begin{algorithm}[!t]
    \small
    \begin{algorithmic}[1]
    \Require {$A_t, S_t, E, n_{chains}$}
    \State $eventChains = E.sample(S_t, n_{chains})$ \Comment{sample chains}
    \State $A = MCTS(A_t,eventChains)$
    \State $\hat{A} \gets \{\}$
    \For {$a \in A$} \Comment{done in parallel}
        \State $\hat{A}.append(mean(a))$ \Comment{aggregate across chains}
    \EndFor
    \State \Return $\argmaxl_{x_i \in X}(\hat{A}[i])$
    \caption{MCTS evaluation}
    \label{algo:mcts_evaluation}
    \end{algorithmic}
\end{algorithm}

\textbf{Sampling Chain Generation:} Running MCTS on a single chain of events would be unable to reflect the uncertainty that can occur in the day-to-day transit operations. We account for this stochasticity by introducing many passenger count event chains either sampled from the generative models or by adding noise to real-world data. 
We use \textit{root parallelization}~\cite{chaslot_parallel_2008} to handle the uncertainty. In root parallelization, a large number of these chains are used to generate multiple trees in parallel at every decision epoch. The best action is decided by the node which has the highest average score across all trees. The process for evaluating and selecting an action is provided in~\cref{algo:mcts_evaluation}. Given the current state at time $t$ and the generative model $E$, the algorithm samples $n$ event chains. A tree is then instantiated and MCTS is executed in parallel (line 2). This returns a matrix $A$ of feasible actions as rows and event chains as columns. We average each row to get the action score over multiple event chains. Finally, the action with the maximum score in $\hat{A}$ is returned (line 7).

\section{Experiments and Results}
\label{sec:experiments_and_results}
We evaluate the performance of the proposed framework's effectiveness on public transit data obtained from a major metropolitan area in the United States. We use real-world data collected and provided by our partner agency.

\subsection{Experimental Setup and Data Description}

\textbf{APC Dataset:} We used real-world APC gathered from the MTA's entire fleet of buses over two years between January 2020 and April 2022. The dataset consists of bus arrival and departure times at every stop along their route, passenger occupancy, and information on whether a particular bus was a regular service bus or an overloaded bus.

\textbf{Travel time and distance matrix:} We generated an empirical distribution of travel times between consecutive pairs of stops present in the dataset. We then used OpenStreetMap and OSMnx to generate distance and travel times between all pairs of stops within the target area. Traffic incidents are factored in using generative models.

\textbf{Generative models:} We used the entire APC dataset along with other spatiotemporal features such as weather, traffic, and holidays, to train predictive 
models for stop-level passenger occupancy. We used the models to generate probabilities for different ranges of possible waiting passengers at the stop. We then selected uniformly at random, the number of boarding passengers from within that range. We generated 40 distinct passenger arrival chains. We used 20 for tuning the hyperparameters and 20 for validating the framework.

\begin{table}[!t]
\scriptsize
\centering
    \caption{Parameter Table}
    \label{tab:hyperparameters}
    \begin{tabular}{|c|c|c|c|}
    \hline
    \textbf{Parameter} & \textbf{Value} & \textbf{Description} & \textbf{Remark} \\ 
    \hline
    \makecell{Planning\\Horizon} & 1 hour & \makecell{Distance in the future\\MCTS will simulate to} & \makecell{problem\\configuration}\\ 
    \hline
    \makecell{Decision\\Epoch Interval} & 0.25 hours & \makecell{Time between forced\\decision-making\\ epochs for agents} & \makecell{problem\\configuration}\\
    \hline
    \makecell{Passenger\\Wait Time} & 0.5 hours & \makecell{Time before\\a passenger\\leaves the stop} & \makecell{problem\\configuration} \\
    \hline
    Discount Factor & 0.99997 & \makecell{Discount factor\\for future rewards} & \makecell{problem\\configuration}\\
    \hline
    \makecell{Number of\\substitute buses} & 5 & \makecell{Number of agents\\considered for\\decision-making} &  \makecell{problem\\configuration} \\ 
    \hline
    \makecell{Event chains} & 20 & \makecell{Accounting for\\stochasticity} &  \makecell{problem\\configuration} \\ 
    \hline
    \makecell{No. of\\MCTS Simulations} & 200  & \makecell{No. of MCTS\\iterations} & hyperparameter \\ 
    \hline
    $\mathcal{C}$ & 1000 & \makecell{Controls exploit\\\& explore characteristics} & hyperparameter \\
    
    \hline
    \end{tabular}
\end{table}

\textbf{System configuration:} First, we constrain the number of stationing stops to 25, instead of every available stop in the environment. This decreases the action space making search tractable while keeping it close to the real world by distributing it evenly across the city. The 25 stops are selected based on discussions with the transit agency. They have flagged these stops as both the current and potential locations where issues often occur. We also select the interval to be 15 minutes between decision epochs. We assume that there are 5 available overload buses to allocate, which is the average number of buses the agency attempts to keep in reserve. Second, there is no limit to the number of disruptions that can happen each day, based on the event chains, there are typically 2-4 disruptions per day. Third, we assume that passengers will arrive at the stop within a 10-minute window before the bus is scheduled to arrive and will wait at the stop for 30 minutes before leaving. A bus that comes early at the stop will wait until the scheduled departure time, while a bus that arrives late will pick up passengers, if there are any, and leave immediately. Finally, we assume that all buses start and end their service at the garage. 
%
We ran experiments on the Chameleon testbed using Intel Xeon Platinum 8380 machines, with 160 cores running at 2.30GHz with 251GB RAM~\cite{keahey2020lessons}.

\textbf{Baseline Greedy Approaches:} We evaluate the performance of our approach against a baseline greedy approach. We used the same number of vehicles, vehicle capacities, incidents, and passenger event chains. According to the agency, they will usually dispatch substitute buses, in a greedy manner if they are available (taking into account domain expertise). Thus, for our greedy baseline, if a bus leaves passengers due to overcrowding, a substitute bus will be dispatched immediately, if the number of people left amounts to 5\% of the last bus' capacity. Also, a substitute bus will be sent immediately, given enough resources, to cover for a broken down bus. In the greedy approaches, there will be no stationing aside from the initial assignment at the start of the service day. There are three possible initial stationing configurations: \textit{GARAGE}, \textit{AGENCY}, and \textit{SEARCH}. GARAGE is equivalent to a no stationing plan, AGENCY represents the current stationing used by the transit office and SEARCH is the result of performing random local search with simulated annealing to find the initial stationing configuration that would serve the most passengers while traveling the least amount of deadhead miles.

\begin{figure}[!t]
    \centering
    \includegraphics[width=0.40\textwidth]{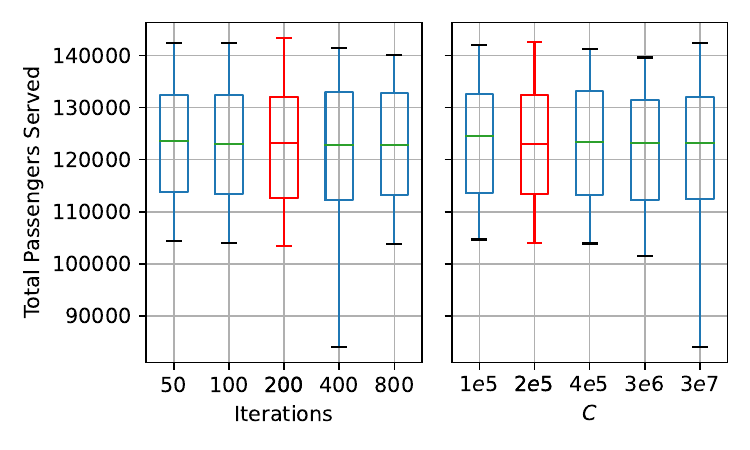}
    \caption{Parameter search for $\mathcal{C}$ and number of MCTS simulations. Selected parameters are colored red, selected due to the highest average passengers served and run time.}
    \label{fig:parameter_tuning}
\end{figure}

\textbf{Parameter tuning:}
We used one week of data for tuning $\mathcal{C}$, the parameter for adjusting exploration over exploitation, and $IT$, the number of MCTS simulations per decision epoch.
This was done both in the interest of time and since public transit follows a weekly schedule resulting in a consistent pattern.
We used 20 of the 40 event chains for training. \cref{fig:parameter_tuning} shows how each parameter affects the performance of the approach over the greedy baseline. The initial value of $C$ was selected by identifying the mean of $\hat{X_j}$ and iteratively decreasing it. We select $\mathcal{C}$ which has the highest improvement over baseline with an acceptable balance of exploration and exploitation. We then select the highest number of MCTS simulations that would still allow us to complete a decision epoch in real time. \cref{fig:decision_time} shows how adjusting this parameter affects the time per decision epoch. The final parameters and configurations for our framework are listed in \cref{tab:hyperparameters}.


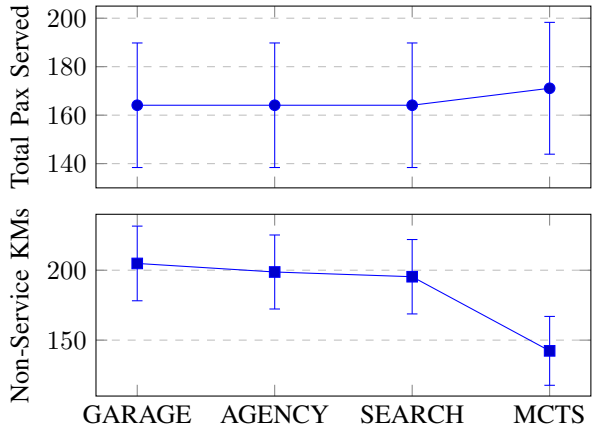
\begin{figure}[!t]
\centering
\begin{tikzpicture}
  \begin{groupplot}[
      group style={
          group size=1 by 2,
          vertical sep=10pt,
          x descriptions at=edge bottom,
      },
      width=0.45\textwidth,
      height=4cm,
      xtick={0, 1,2,3},
      xticklabels={GARAGE, AGENCY, SEARCH, MCTS},
      ymajorgrids=true,
      grid style=dashed,
  ]

  \nextgroupplot[ylabel={Total Pax Served}, ymin=130, ymax=205]
  \addplot+ [
      mark=*, 
      error bars/.cd, 
      y dir=both, 
      y explicit
  ] table [
      x expr=\coordindex, 
      y=mean, 
      y error=std,
      col sep=space
  ] {
      method std mean type
      BASELINE 25.694793 164.090909 total_served
      AGENCY 25.694793 164.090909 total_served
      SEARCH 25.694793 164.090909 total_served
      MCTS 27.195483 171.090909 total_served
  };

  \nextgroupplot[ylabel={Non-Service KMs}, ymin=110, ymax=240]
  \addplot+ [
      mark=square*, 
      error bars/.cd, 
      y dir=both, 
      y explicit
  ] table [
      x expr=\coordindex, 
      y=mean, 
      y error=std,
      col sep=space
  ] {
      method std mean type
      BASELINE 26.698496 204.841818 service_kms
      AGENCY 26.466707 198.724545 service_kms
      SEARCH 26.603135 195.343636 service_kms
      MCTS 24.627822 142.323636 service_kms
  };
  
  \end{groupplot}
\end{tikzpicture}

\caption{Comparing the performance of our approach to baselines when both the real world environment and MDP environment are sampled from generative models.}
\label{fig:model_environment_model_chains}
\end{figure}


\subsection{Experimental Scenarios}


We ran experiments on two weeks of data unseen during hyperparameter tuning and compared the baseline policies with our proposed approach. We ran two sets of experiments, shown in \cref{tab:experiment_scenarios}. First, we used a generative model to represent the real-world environment and used event chains sampled from the generative model as the MCTS environment. Next, we validated our approach with an environment based on actual empirical data and chains based on real-world passenger counts with Gaussian noise applied to it and the models sampled from the generative model. The passenger arrival chains sampled from the generative models have a much larger passenger count, due to how the generative model was trained. 


\subsection{Results}

The results for total passengers served are shown in \cref{fig:model_environment_model_chains}. The first observation is that by using the proposed framework, we improve the total number of passengers served by \textbf{~$10$} passengers on average. This number can increase further when there are more instances of overage events and fewer disruption events. Since a bus that has been assigned to take over a broken bus is essentially unable to be dispatched to overage events, it reduces the window for optimization.


We also observe a significant reduction in total deadhead miles traveled by the overload buses when we perform dispatch using the proposed approach. We save on average around \textbf{$50$} kilometers per overload bus. This reduction in deadhead miles seen in \cref{fig:model_environment_model_chains}, is not unexpected. This can be attributed to the fact that our approach does not dispatch buses when the estimated reward is minimal, which leaves it available for future events with a larger reward. 



The ability to make non-myopic decisions for when and where to allocate and dispatch overflow buses results in our approach performing better than the baseline. Consider a major overage at some time in the future. The baseline will continuously dispatch buses for minor overages before this moment, effectively exhausting its resources and leaving it unable to serve the major incident. These scenarios are effectively handled by our framework. 

\begin{table}[!t]

\definecolor{Gray}{gray}{0.9}
\scriptsize
\centering
    \caption{Experimental Scenarios}
    \label{tab:experiment_scenarios}
    \begin{tabular}{|l|cccc|}
    \hline
    \textbf{Scenario} & \textbf{Env.} & \textbf{Chains} & \textbf{Traffic} & \textbf{Breakdown} \\     
    \hline
    Full generative & Generative & Generative & True & True\\   \hline\rowcolor{Gray}
    \hline
    Noisy Real & Real-world & Noisy real & True & True\\
    \hline
    Generative Model & Real-world & Generative & True & True\\
    \hline
    \end{tabular}
\end{table}


\begin{figure}[!t]
\centering

\begin{tikzpicture}
\begin{axis}[
    height=5cm,
    width=0.49\textwidth,
    ylabel={Non-Service KMs},
    xtick={0,...,4}, 
    xticklabels={GARAGE, AGENCY, SEARCH, MCTS, MCTS\_5},
    x tick label style={
        rotate=0, 
        font=\fontsize{8}{10}\selectfont 
    },
    grid=major,
    legend pos=north west,
    error bars/y dir=both,
    error bars/y explicit,
]

\addplot+[error bars/.cd, y dir=both, y explicit] 
table[x expr=\coordindex, y=mean, y error=std, col sep=comma] {data.csv};

\addplot+[mark=*, only marks, forget plot] 
table[x expr=\coordindex, y=mean, col sep=comma] {data.csv};

\end{axis}
\end{tikzpicture}
\caption{Comparing deadhead miles of the approach against the different stationing baseline on the real-world data. MCTS\_5 is the same approach except decisions are done at 5-minute intervals compared to 15-minute intervals.}
\label{fig:realworld_chains}
\end{figure}
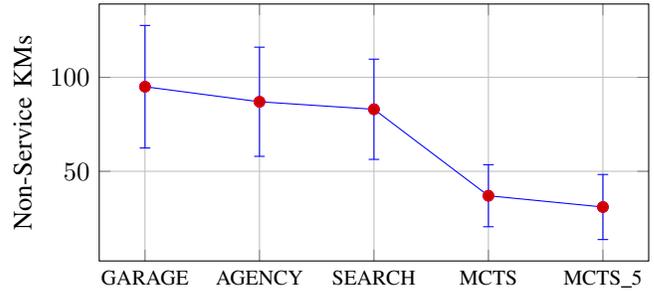

\begin{figure}[!t]
\centering
\includegraphics[width=0.40\textwidth]{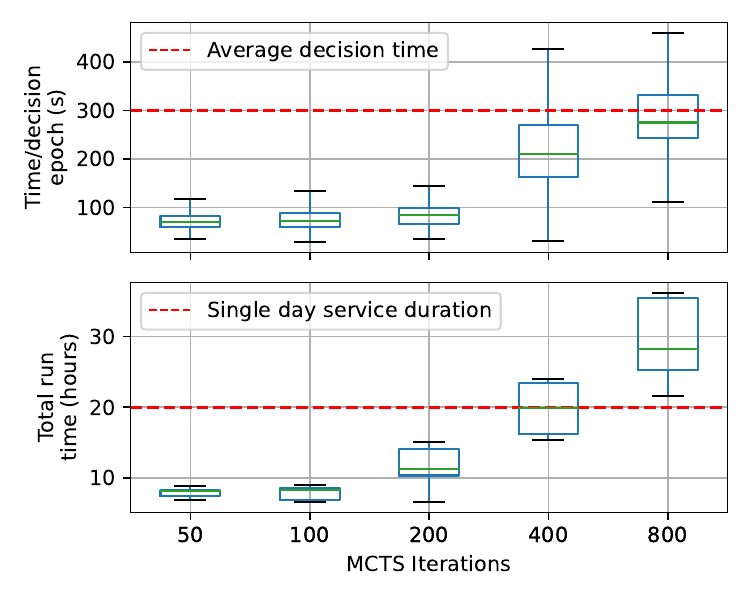}
\caption{Measuring processing time per decision epoch and overall run time for an entire service day.}
\label{fig:decision_time}
\end{figure}

\textbf{Real-world effectiveness}: The end goal of our work is to be integrated into a multi-purpose tool that public transit agencies, for any major metropolitan city, can use to improve their system. Thus, we evaluate the effectiveness of our framework in a real-world environment. We use empirical data as the real-world environment and sample chains from generative models for the simulation environment. Since the passenger overages and disruption events are very sparse in the real-world data, both the baselines and our MCTS approach can serve the maximum feasible amount of passengers that can be served. However, \cref{fig:realworld_chains} shows that our approach can outperform all baselines in terms of total deadhead miles traveled by substitute buses. This is to be expected, the strength of our approach lies in how well we can estimate and simulate the conditions of the real world using generative models. This savings in distance traveled translates to savings in manpower and resources for the agency.


\textbf{Computation Time:} Improving upon their stationing and dispatch methodology is only realistic if our online approach can produce results within the time that they must make their decisions. \cref{fig:decision_time} shows how adjusting the number of MCTS simulations affects both processing time decision epoch and overall run time. The figure above shows that keeping the number of MCTS simulations below $400$ will ensure that decision epochs finish at or below the five-minute interval that MTA uses as their decision intervals for their decision-making. The figure below shows the total run time in real-time for our approach to completely simulate the entire service day and provide optimal actions at each decision epoch.  
In our experiments, the entire service day lasts on average around 20 hours (04:00 am to 24:00 am). Thus, running $200$ MCTS simulations per chain lets us finish all simulations and decision-making in $20$ hours. 

However, it should be noted that the total run times and computation costs are also subject to several other parameters shown in~\cref{tab:hyperparameters}. Increasing the number of event chains may make the framework adapt to randomness in the environment, but also increase computation. Increasing the number of simulations may allow the MCTS to converge to a more optimal solution while increasing run times. Our approach is largely bounded by CPU, with minimal memory and disk requirements during run time.
The desire to be able to perform in real-time for actual deployment is one of the reasons we selected certain hyperparameters such as the number of MCTS simulations and chains. Based on our results, 20 chains is an adequate lower bound for the number of event chains. Each chain takes one CPU core and is limited by the CPU clock speeds. Also, 200 MCTS simulations provide acceptable results while keeping the decision epoch below the decision-making time suggested by the partner agency. 
For larger cities, with different transit requirements, our approach may be able to perform better by varying different hyperparameters.




\section{Related Work}
\label{sec:related_work}
Scheduling for fixed-line bus systems can be divided into static and dynamic scheduling.
Static scheduling is the process of designing the fixed routes beforehand and is an offline problem.
This includes the network route design, setting timetables, and scheduling vehicles to service routes.
In this context, static scheduling for fixed-line transit is a version of the classic line planning problem \cite{schobel2012line}, \cite{borndorfer2007column}, \cite{gkiotsalitis2021public}.
The design of a static fixed-line bus system aims to meet various, often competing, objectives including coverage, meeting demand, serviceability requirements, and cost.
Recent work has used MCTS to efficiently evaluate and search potential transit routes based on multiple objectives \cite{weng2020pareto}.

We assume that the fixed-line schedules are designed beforehand and are outside the scope of this work.
We focus on dynamic scheduling where the goal is to adapt in real-time to address dynamic demand and uncertainty in the system.
Current dynamic scheduling for fixed-line services is largely devoted to studying bus holding strategies aimed at improving bus service from the passenger's perspective.
This involves designing a control system that manages the arrival times of vehicles along a route to minimize headway, passenger waiting time, or other serviceability metrics.
The control strategy includes: holding control \cite{eberlein2001holding} \cite{wu2016designing} \cite{wang2020proactive}, bus speed control \cite{daganzo2011reducing} \cite{wang2020proactive}, stop skipping \cite{wang2020proactive} and boarding limits.

Liu et al extend the notion of control strategies to focus on dynamic dispatching \cite{liu2021robust}.
Their work focuses on dispatching vehicles for a single bus line.
They monitor current demand, forecasted demand, and headway information along the line and then schedule when the next vehicle will leave the depot.
Vehicles are not allowed to be dispatched to any stop or any trip, only at the first stop along the trip in question.
Unlike our work, Liu et al do not consider reallocation strategies.
Reallocation strategies have proven highly effective in rideshare and ride pooling applications \cite{alonso2017demand, lin2018efficient, tang2021value}.
The success of reallocation strategies in rideshare and ride pooling motivates our use of reallocation and stationing in the context of dynamic fixed-line transit.

MCTS has proven effective for planning in a variety of domains \cite{swiechowski2022monte}.
Planning with MCTS typically takes one of two forms.
In the first setting, a value or state-action function is learned offline through reinforcement learning and can be queried at inference time to help guide the search process \cite{silver2016mastering}.
The second setting is purely online.
In purely online MCTS generative models of the environment are used at inference time to evaluate future actions in the context of expected demand \cite{claes2017decentralised, mukhopadhyay2019online, pettet2021hierarchical}.
Pure online approaches have the benefit of not relying on value functions learned offline that can become stale in rapidly changing environments such as urban systems \cite{wilbur2022online}.
Therefore, this work focuses on the second approach where we use MCTS with passenger distribution and traffic incident generative models based on historical data.

\section{Conclusion}
\label{sec:conclusion}
We design a non-myopic, always online approach for optimizing stationing and dispatch of reserve or overflow buses for fixed-line transit that is scalable to real-world applications. 
We introduce several approximations and assumptions to allow our approach to limit the exponential action space present in the real world.
We demonstrated that our approach improves upon the baseline by 2\% on average, and as much as 7\% on total passengers served. An additional benefit is that our approach reduces the overall deadhead miles traveled by overflow buses by 42\% on average, as they are stationed and dispatched across the city. We also show that our approach, along with the generative models, can adapt to the real-world environment. Thus, performing similarly to event chains based on the real world. Finally, as the future of our implementation is to be deployed as a tool for public transit agencies, we show that with careful tuning of parameters, we can execute decision epochs within the cut-off period for their decision-making.

\section*{Acknowledgment}
This material is based upon work sponsored by the National Science Foundation under Award Number 1952011 and the Federal Transit Administration COVID-19 Research Grant under Federal Award Identification Number TN-2021-015-00. Results presented in this paper were obtained using the Chameleon Testbed supported by the NSF.
 
\bibliographystyle{IEEEtran}
\bibliography{references,extra}

\begin{thebibliography}{10}
\providecommand{\url}[1]{#1}
\csname url@samestyle\endcsname
\providecommand{\newblock}{\relax}
\providecommand{\bibinfo}[2]{#2}
\providecommand{\BIBentrySTDinterwordspacing}{\spaceskip=0pt\relax}
\providecommand{\BIBentryALTinterwordstretchfactor}{4}
\providecommand{\BIBentryALTinterwordspacing}{\spaceskip=\fontdimen2\font plus
\BIBentryALTinterwordstretchfactor\fontdimen3\font minus \fontdimen4\font\relax}
\providecommand{\BIBforeignlanguage}[2]{{%
\expandafter\ifx\csname l@#1\endcsname\relax
\typeout{** WARNING: IEEEtran.bst: No hyphenation pattern has been}%
\typeout{** loaded for the language `#1'. Using the pattern for}%
\typeout{** the default language instead.}%
\else
\language=\csname l@#1\endcsname
\fi
#2}}
\providecommand{\BIBdecl}{\relax}
\BIBdecl

\bibitem{status_highways_2003}
\BIBentryALTinterwordspacing
F.~H.~A. U.S. Department~of Transportation, ``Status of the nation’s highways, bridges, and transit: 2002 conditions and performance report to congress,'' 2003. [Online]. Available: \url{https://www.apta.com/wp-content/uploads/2022-Q2-Ridership-APTA.pdf}
\BIBentrySTDinterwordspacing

\bibitem{ridership_report_2022}
\BIBentryALTinterwordspacing
A.~P.~T. Association, ``Public transportation ridership report,'' 2022. [Online]. Available: \url{https://www.apta.com/wp-content/uploads/2022-Q2-Ridership-APTA.pdf}
\BIBentrySTDinterwordspacing

\bibitem{fu_real-time_2003}
\BIBentryALTinterwordspacing
L.~Fu, Q.~Liu, and P.~Calamai, ``\BIBforeignlanguage{en}{Real-{Time} {Optimization} {Model} for {Dynamic} {Scheduling} of {Transit} {Operations}},'' \emph{\BIBforeignlanguage{en}{Transportation Research Record: Journal of the Transportation Research Board}}, vol. 1857, no.~1, pp. 48--55, Jan. 2003. [Online]. Available: \url{http://journals.sagepub.com/doi/10.3141/1857-06}
\BIBentrySTDinterwordspacing

\bibitem{Kochenderfer_2015}
M.~J. Kochenderfer, C.~Amato, G.~Chowdhary, J.~P. How, H.~J.~D. Reynolds, J.~R. Thornton, P.~A. Torres-Carrasquillo, N.~K. \"{U}re, and J.~Vian, \emph{Decision Making Under Uncertainty: Theory and Application}, 1st~ed.\hskip 1em plus 0.5em minus 0.4em\relax The MIT Press, 2015.

\bibitem{jung_dynamic_2016}
J.~Jung, R.~Jayakrishnan, and J.~Y. Park, ``Dynamic {Shared}-{Taxi} {Dispatch} {Algorithm} with {Hybrid}-{Simulated} {Annealing},'' \emph{Computer-Aided Civil and Infrastructure Engineering}, vol.~31, no.~4, pp. 275--291, 2016.

\bibitem{newell_1971}
\BIBentryALTinterwordspacing
G.~F. Newell, ``Dispatching policies for a transportation route,'' \emph{Transportation Science}, vol.~5, no.~1, pp. 91--105, 1971. [Online]. Available: \url{http://www.jstor.org/stable/25767595}
\BIBentrySTDinterwordspacing

\bibitem{hu2007markov}
Q.~Hu and W.~Yue, \emph{Markov decision processes with their applications}.\hskip 1em plus 0.5em minus 0.4em\relax Springer Science \& Business Media, 2007, vol.~14.

\bibitem{MA20131}
\BIBentryALTinterwordspacing
X.~Ma, Y.-J. Wu, Y.~Wang, F.~Chen, and J.~Liu, ``Mining smart card data for transit riders’ travel patterns,'' \emph{Transportation Research Part C: Emerging Technologies}, vol.~36, pp. 1--12, 2013. [Online]. Available: \url{https://www.sciencedirect.com/science/article/pii/S0968090X13001630}
\BIBentrySTDinterwordspacing

\bibitem{apc_data}
\BIBentryALTinterwordspacing
J.~P. Talusan, A.~Mukhopadhyay, D.~Freudberg, and A.~Dubey, ``On designing day ahead and same day ridership level prediction models for city-scale transit networks using noisy apc data,'' in \emph{2022 IEEE International Conference on Big Data (Big Data)}.\hskip 1em plus 0.5em minus 0.4em\relax Los Alamitos, CA, USA: IEEE Computer Society, Dec. 2022, pp. 5598--5606. [Online]. Available: \url{https://doi.ieeecomputersociety.org/10.1109/BigData55660.2022.10020390}
\BIBentrySTDinterwordspacing

\bibitem{browne_survey_2012}
\BIBentryALTinterwordspacing
C.~B. Browne, E.~Powley, D.~Whitehouse, S.~M. Lucas, P.~I. Cowling, P.~Rohlfshagen, S.~Tavener, D.~Perez, S.~Samothrakis, and S.~Colton, ``\BIBforeignlanguage{en}{A {Survey} of {Monte} {Carlo} {Tree} {Search} {Methods}},'' \emph{\BIBforeignlanguage{en}{IEEE Transactions on Computational Intelligence and AI in Games}}, vol.~4, no.~1, pp. 1--43, Mar. 2012. [Online]. Available: \url{http://ieeexplore.ieee.org/document/6145622/}
\BIBentrySTDinterwordspacing

\bibitem{kocsis_bandit_2006}
L.~Kocsis and C.~Szepesvári, ``Bandit {Based} {Monte}-{Carlo} {Planning},'' in \emph{Machine {Learning}: {ECML} 2006}, J.~Fürnkranz, T.~Scheffer, and M.~Spiliopoulou, Eds.\hskip 1em plus 0.5em minus 0.4em\relax Berlin, Heidelberg: Springer Berlin Heidelberg, 2006, pp. 282--293.

\bibitem{chaslot_parallel_2008}
G.~M. J.~B. Chaslot, M.~H.~M. Winands, and H.~J. van~den Herik, ``Parallel {Monte}-{Carlo} {Tree} {Search},'' in \emph{Computers and {Games}}, H.~J. van~den Herik, X.~Xu, Z.~Ma, and M.~H.~M. Winands, Eds.\hskip 1em plus 0.5em minus 0.4em\relax Berlin, Heidelberg: Springer Berlin Heidelberg, 2008, pp. 60--71.

\bibitem{keahey2020lessons}
K.~Keahey, J.~Anderson, Z.~Zhen, P.~Riteau, P.~Ruth, D.~Stanzione, M.~Cevik, J.~Colleran, H.~S. Gunawi, C.~Hammock, J.~Mambretti, A.~Barnes, F.~Halbach, A.~Rocha, and J.~Stubbs, ``Lessons learned from the chameleon testbed,'' in \emph{Proceedings of the 2020 USENIX Annual Technical Conference (USENIX ATC '20)}.\hskip 1em plus 0.5em minus 0.4em\relax USENIX Association, July 2020.

\bibitem{schobel2012line}
A.~Sch{\"o}bel, ``Line planning in public transportation: models and methods,'' \emph{OR spectrum}, vol.~34, no.~3, pp. 491--510, 2012.

\bibitem{borndorfer2007column}
R.~Bornd{\"o}rfer, M.~Gr{\"o}tschel, and M.~E. Pfetsch, ``A column-generation approach to line planning in public transport,'' \emph{Transportation Science}, vol.~41, no.~1, pp. 123--132, 2007.

\bibitem{gkiotsalitis2021public}
K.~Gkiotsalitis and O.~Cats, ``Public transport planning adaption under the covid-19 pandemic crisis: literature review of research needs and directions,'' \emph{Transport Reviews}, vol.~41, no.~3, pp. 374--392, 2021.

\bibitem{weng2020pareto}
D.~Weng, R.~Chen, J.~Zhang, J.~Bao, Y.~Zheng, and Y.~Wu, ``Pareto-optimal transit route planning with multi-objective monte-carlo tree search,'' \emph{IEEE Transactions on Intelligent Transportation Systems}, vol.~22, no.~2, pp. 1185--1195, 2020.

\bibitem{eberlein2001holding}
X.~J. Eberlein, N.~H. Wilson, and D.~Bernstein, ``The holding problem with real--time information available,'' \emph{Transportation science}, vol.~35, no.~1, pp. 1--18, 2001.

\bibitem{wu2016designing}
W.~Wu, R.~Liu, and W.~Jin, ``Designing robust schedule coordination scheme for transit networks with safety control margins,'' \emph{Transportation Research Part B: Methodological}, vol.~93, pp. 495--519, 2016.

\bibitem{wang2020proactive}
W.~Wang, F.~Zong, and B.~Yao, ``A proactive real-time control strategy based on data-driven transit demand prediction,'' \emph{IEEE Transactions on Intelligent Transportation Systems}, vol.~22, no.~4, pp. 2404--2416, 2020.

\bibitem{daganzo2011reducing}
C.~F. Daganzo and J.~Pilachowski, ``Reducing bunching with bus-to-bus cooperation,'' \emph{Transportation Research Part B: Methodological}, vol.~45, no.~1, pp. 267--277, 2011.

\bibitem{liu2021robust}
Y.~Liu, X.~Luo, X.~Wei, Y.~Yu, and J.~Tang, ``Robust optimization model for single line dynamic bus dispatching,'' \emph{Sustainability}, vol.~14, no.~1, p.~73, 2021.

\bibitem{alonso2017demand}
J.~Alonso-Mora, S.~Samaranayake, A.~Wallar, E.~Frazzoli, and D.~Rus, ``On-demand high-capacity ride-sharing via dynamic trip-vehicle assignment,'' \emph{Proceedings of the National Academy of Sciences}, vol. 114, no.~3, pp. 462--467, 2017.

\bibitem{lin2018efficient}
K.~Lin, R.~Zhao, Z.~Xu, and J.~Zhou, ``Efficient large-scale fleet management via multi-agent deep reinforcement learning,'' in \emph{Proceedings of the 24th ACM SIGKDD International Conference on Knowledge Discovery \& Data Mining}, 2018, pp. 1774--1783.

\bibitem{tang2021value}
X.~Tang, F.~Zhang, Z.~Qin, Y.~Wang, D.~Shi, B.~Song, Y.~Tong, H.~Zhu, and J.~Ye, ``Value function is all you need: A unified learning framework for ride hailing platforms,'' in \emph{Proceedings of the 27th ACM SIGKDD Conference on Knowledge Discovery \& Data Mining}, 2021, pp. 3605--3615.

\bibitem{swiechowski2022monte}
M.~{\'S}wiechowski, K.~Godlewski, B.~Sawicki, and J.~Ma{\'n}dziuk, ``Monte carlo tree search: A review of recent modifications and applications,'' \emph{Artificial Intelligence Review}, pp. 1--66, 2022.

\bibitem{silver2016mastering}
D.~Silver, A.~Huang, C.~J. Maddison, A.~Guez, L.~Sifre, G.~Van Den~Driessche, J.~Schrittwieser, I.~Antonoglou, V.~Panneershelvam, M.~Lanctot \emph{et~al.}, ``Mastering the game of go with deep neural networks and tree search,'' \emph{nature}, vol. 529, no. 7587, pp. 484--489, 2016.

\bibitem{claes2017decentralised}
D.~Claes, F.~Oliehoek, H.~Baier, and K.~Tuyls, ``Decentralised online planning for multi-robot warehouse commissioning,'' in \emph{Proceedings of the 16th Conference on Autonomous Agents and MultiAgent Systems}, ser. AAMAS '17.\hskip 1em plus 0.5em minus 0.4em\relax Richland, SC: International Foundation for Autonomous Agents and Multiagent Systems, 2017, p. 492–500.

\bibitem{mukhopadhyay2019online}
A.~Mukhopadhyay, G.~Pettet, C.~Samal, A.~Dubey, and Y.~Vorobeychik, ``An online decision-theoretic pipeline for responder dispatch,'' in \emph{Proceedings of the 10th ACM/IEEE International Conference on Cyber-Physical Systems}, 2019, pp. 185--196.

\bibitem{pettet2021hierarchical}
\BIBentryALTinterwordspacing
G.~Pettet, A.~Mukhopadhyay, M.~J. Kochenderfer, and A.~Dubey, ``Hierarchical planning for dynamic resource allocation in smart and connected communities,'' \emph{ACM Trans. Cyber-Phys. Syst.}, vol.~6, no.~4, nov 2022. [Online]. Available: \url{https://doi.org/10.1145/3502869}
\BIBentrySTDinterwordspacing

\bibitem{wilbur2022online}
\BIBentryALTinterwordspacing
M.~Wilbur, S.~Kadir, Y.~Kim, G.~Pettet, A.~Mukhopadhyay, P.~Pugliese, S.~Samaranayake, A.~Laszka, and A.~Dubey, ``An online approach to solve the dynamic vehicle routing problem with stochastic trip requests for paratransit services,'' in \emph{2022 ACM/IEEE 13th International Conference on Cyber-Physical Systems (ICCPS)}.\hskip 1em plus 0.5em minus 0.4em\relax Los Alamitos, CA, USA: IEEE Computer Society, may 2022, pp. 147--158. [Online]. Available: \url{https://doi.ieeecomputersociety.org/10.1109/ICCPS54341.2022.00020}
\BIBentrySTDinterwordspacing

\end{thebibliography}

\appendices
\section*{Appendix}

\subsection{Notation Lookup}
\label{appendix:notation_lookup}

\definecolor{Gray}{gray}{0.9}
\begin{table}[H]
\centering
\caption{Notation Lookup Table}
\footnotesize
\begin{tabular}{|l|l|}
\hline

{\textbf{Symbol}}     & {\textbf{Definition}}                           \\ \hline\rowcolor{Gray}
$T$                 	  & Set of trips $i$                                    \\ \hline
$L$                       & Set of stops                                        \\ \hline\rowcolor{Gray}
$e(L^i_j)$                & Scheduled arrival time for stop $j$ on trip $i$     \\ \hline
$V$                       & Set of all buses                                    \\ \hline\rowcolor{Gray}
$V^R$                     & Set of regular buses                                \\ \hline
$V^O$                     & Set of overload buses                               \\ \hline\rowcolor{Gray}
$S$                       & Set of states                                       \\ \hline
$A$                       & Set of all feasible actions                         \\ \hline\rowcolor{Gray}
$P$                  	  & State transition function                           \\ \hline
$T$                       & Temporal distribution over transitions              \\ \hline\rowcolor{Gray}
$R$                       & Instantaneous reward function                       \\ \hline
$\gamma$                  & Discount factor for future rewards                  \\ \hline\rowcolor{Gray}
$vloc_t$                  & Vehicle location                                    \\ \hline
$c(V^i)$                  & Capacity of vehicle                                 \\ \hline\rowcolor{Gray}
$o(V^i_t)$                & Occupancy of vehicle at time $t$                    \\ \hline
$b(L^i_j)$                & Number of boarding for stop $j$ on trip $i$         \\ \hline\rowcolor{Gray}
$w(V^i_t)$                & Status of vehicle at time $t$                       \\ \hline
$C$                       & Explore and exploit parameter                       \\ \hline\rowcolor{Gray}
$E$                       & Generative model                       \\ \hline
\end{tabular}%
\label{tab:symbols}
\end{table}

\subsection{Dispatch actions}
\label{appendix:dispatch_actions}

Our approach improves on the baseline by performing non-myopic decisions for when and where to dispatch overflow buses. Fig.~\ref{fig:response_timeseries_timeline} shows how our approach can more strategically dispatch overflow buses in anticipation of a more significant number of passengers in the future. The greedy baseline (bottom half) exhausts all available overflow buses at around the 19:40 mark, serving fewer passengers. This results in a later trip that starts around 19:45 with a large number of passengers waiting to board, overcrowded and without any overload buses to assist. This results in fewer passengers being served. Overall, our approach (top half) can serve 18\% more passengers at the same time as the baseline. This highlights the shortcomings of myopic approaches which immediately dispatch buses on the first sign of overages. This causes ``stuttering" which also affects the total deadhead miles traveled by overload buses as they keep getting dispatched to trips with minimal rewards.

\begin{figure}[htpb]
    \centering
    {
        \includegraphics[width=0.9\linewidth]{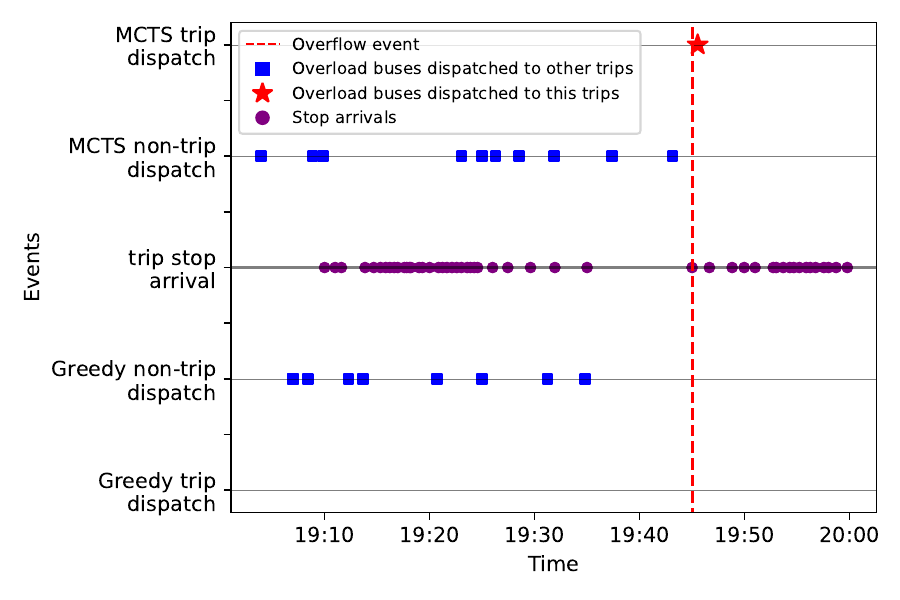}
        \caption{Series of events leading to the dispatch of overflow buses, a bus is dispatched for an overflow event when using our approach, which is absent in the baseline.}
        \label{fig:response_timeseries_timeline}
    }\hfill
\end{figure}

The results are shown in figure~\ref{fig:disruption_response_timeline}. We can observe that the greedy baseline approach dispatches a substitute bus almost immediately after the disruption occurs. However, there have also been instances where a substitute bus is only available, long after the current disrupted trip's schedule has elapsed, stranding people. Meanwhile, the proposed approach estimates that there are no rewards if it decides to send the bus early, instead choosing to wait before sending a substitute bus. While minimal in this example, the results lead to an improvement in total passengers served.

\begin{figure}[t]
    \centering
    {
        \includegraphics[width=0.9\linewidth]{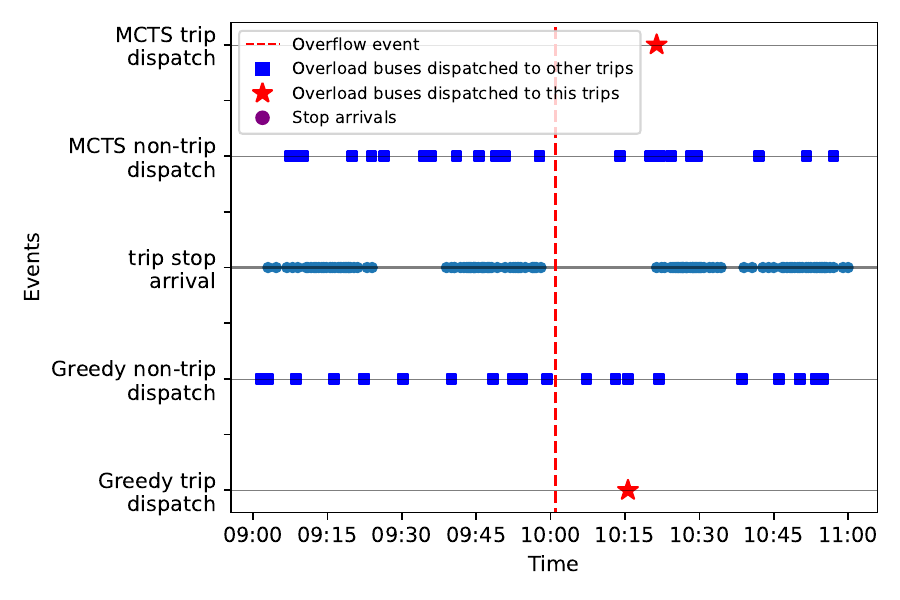}
        \caption{Series of events leading to the dispatch of overflow buses, a disruption event occurs at 10:00 am and both approaches dispatch a bus after some time.}
        \label{fig:disruption_response_timeline}
    }\hfill
\end{figure}

\subsection{Artifact Evaluation}


\begin{table}[ht]
    \centering
    \begin{tabularx}{\linewidth}{|l|X|}
        \hline
        \rowcolor[HTML]{EFEFEF} 
        \textbf{Component} & \textbf{Specification} \\
        \hline
        Programs & Docker \\
        \hline
        RAM & 94GB total, 300GB swap \\
        \hline
        CPU & At least 4 cores \\
        \hline
        OS & Ubuntu, at least 18.04.5 LTS \\
        \hline
    \end{tabularx}
    \caption{Minimum Requirements}
    \label{tab:minimum-reqs}
\end{table}

The included Repeatability Evaluation Package (REP) will cover 
~\cref{fig:model_environment_model_chains}, 
~\cref{fig:realworld_chains}, 
and 
~\cref{fig:decision_time}. However, in the interest of time, this REP will only run on a fraction of the datasets used in the paper. This REP will not include hyperparameter search components, namely ~\cref{fig:parameter_tuning} and ~\cref{tab:hyperparameters}. The results are still valid and proper, and follow a similar trend to the components in the paper. With the current changes and using a machine that meets the minimum requirements~\cref{tab:minimum-reqs}, this REP will take around 4 hours to complete.
Running and generating the figures that exactly match what is in the paper requires at least a CPU with 10 cores and almost 100GB of memory and will take around a few days without parallelization. We have included clear instructions on configuring the current REP to run hyperparameter searches and potentially extend our work for the future.

\subsection*{Installation and Run Instructions}
Instructions are for UNIX-based machines. The link provided below provides additional instructions for Windows.

\begin{enumerate}[nosep]
    \item Download both ``REP\_57.tar.gz'' and ``ARTIFACT\_FILES.tar.gz'' from \url{https://zenodo.org/records/10594255}
    \item Extract the ``REP\_57.tar.gz'' and move the ``ARTIFACT\_FILES.tar.gz'' into the REP\_57 folder:
\begin{verbatim}
tar -xzvf REP_57.tar.gz
mv ARTIFACT_FILES.tar.gz REP_57
cd REP_57
\end{verbatim}

    \item Extract data files using:
\begin{verbatim}
tar -xzvf ARTIFACT_FILES.tar.gz
\end{verbatim}

    \item Build the Docker image, do not forget the period at the end of the command.
\begin{verbatim}
docker build -t iccps2024_stationing .
\end{verbatim}

    \item Run the experiment~\footnote{REP. Note: this can run in the background using ``-d" argument.}: 
\begin{verbatim}
docker run -d -v \
$PWD/code_root:/usr/src/app/code_root \
iccps2024_stationing
\end{verbatim}

    \item Logs during the run can be monitored using the Docker ID generated in the previous step.
\begin{verbatim}
docker logs -f DOCKER_ID
\end{verbatim}

    \item Plots will appear in:
\begin{verbatim}
code_root\experiments\TEST\plots
\end{verbatim}

    \item Raw logs will appear in two locations:
\begin{verbatim}
code_root\experiments\TEST\results
code_root\experiments\TEST\logs
\end{verbatim}

\end{enumerate}

\subsection*{REP Output}
The following figures will be generated by the REP. While this does not directly match the components presented in the paper, they follow a similar trend. Given the computational requirements and time constraints, the figures are truncated versions of those presented in the paper.

\begin{figure}[htpb]
    \centering
    {
        \includegraphics[width=0.7\linewidth]{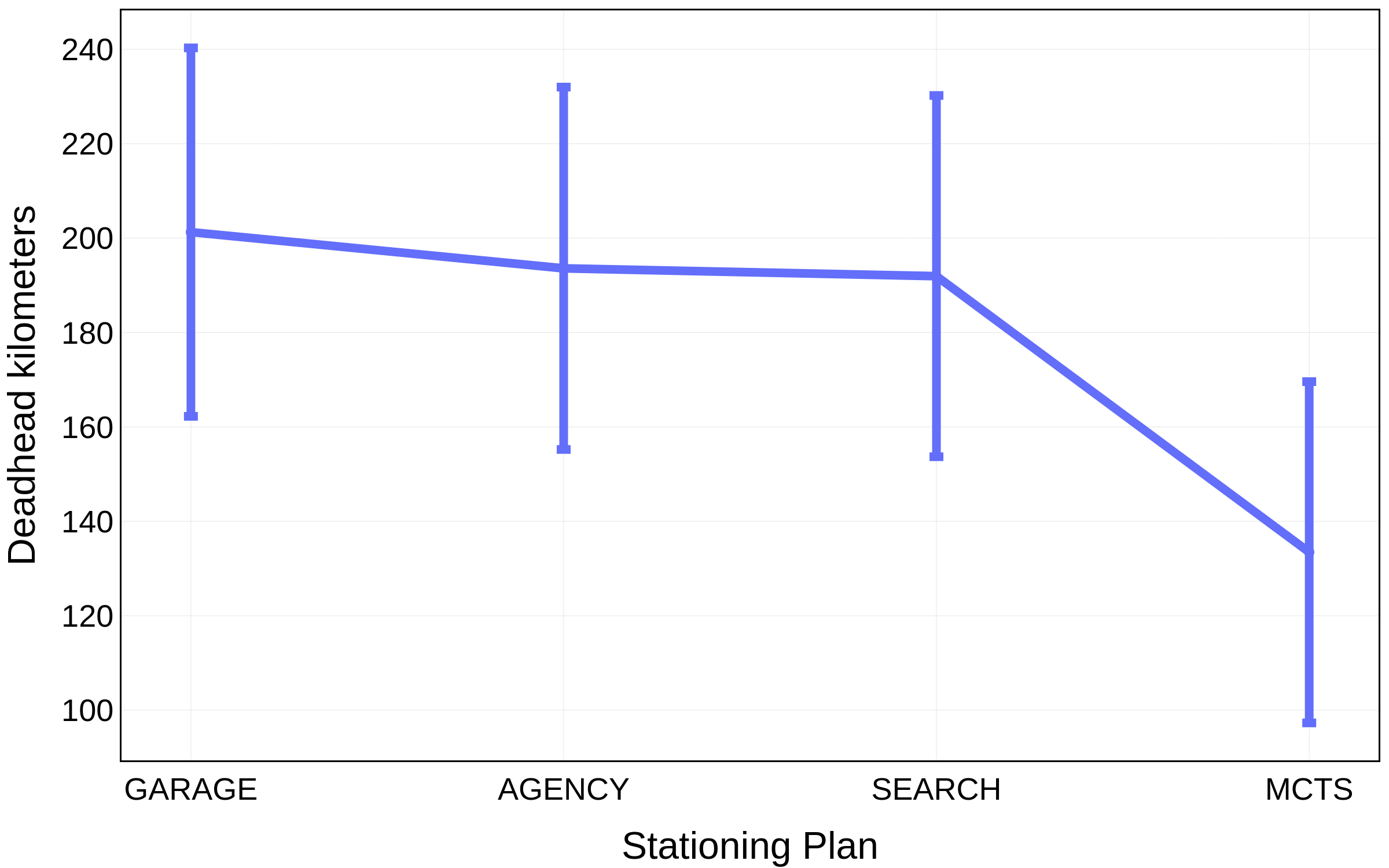}
        \caption{REP output for the bottom portion of ~\cref{fig:model_environment_model_chains}.}
        \label{fig:artifact_fig4b}
    }\hfill
\end{figure}

\begin{figure}[htpb]
    \centering
    {
        \includegraphics[width=0.7\linewidth]{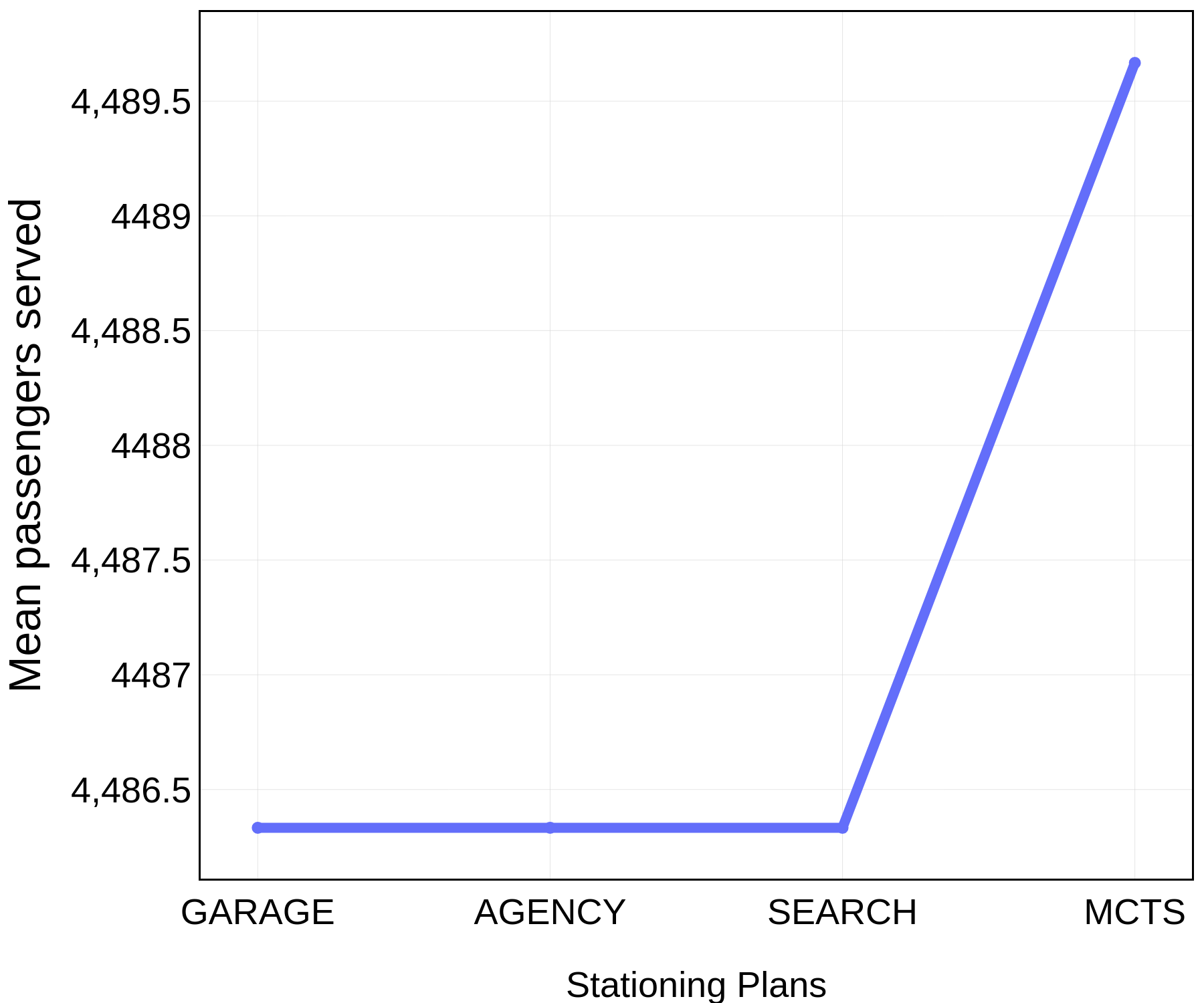}
        \caption{REP output for the top portion of ~\cref{fig:model_environment_model_chains}.}
        \label{fig:artifact_fig4t}
    }\hfill
\end{figure}


\begin{figure}[htpb]
    \centering
    {
        \includegraphics[width=0.7\linewidth]{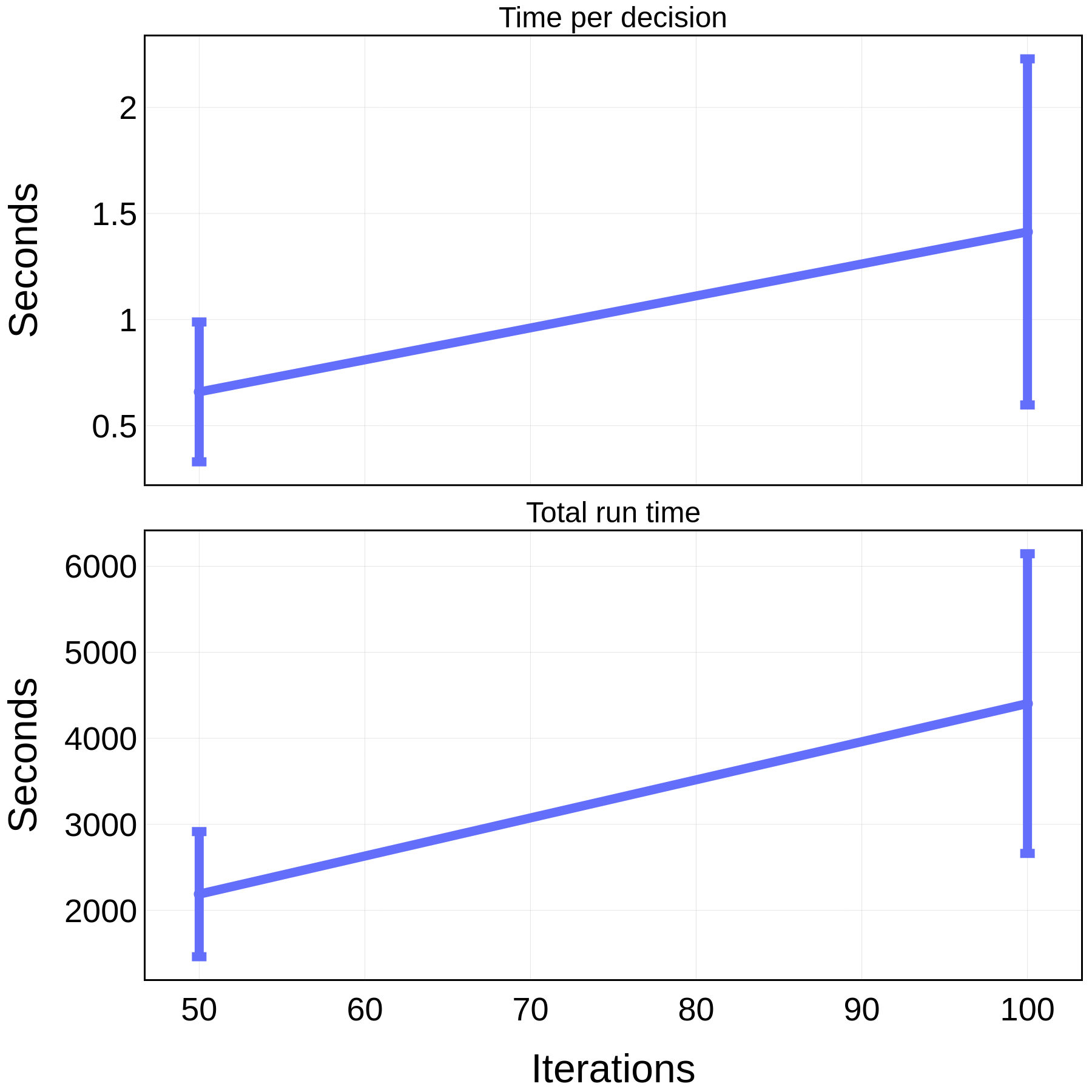}
        \caption{REP output for ~\cref{fig:decision_time}.}
        \label{fig:artifact_fig7}
    }\hfill
\end{figure}

\subsection*{Troubleshooting}
Here are some issues commonly encountered when using Docker and running the REP.

\begin{enumerate}
  \item If running using Docker Desktop on a Mac, you might need to allow file sharing on the current git repo directory.
  \item The time in the results is in UTC. The first one contains the raw logs detailing the bus movement and passenger pickups and dropoffs. The second one is a summary containing 3 distinct CSVs and a summary of the results at the bottom.
  \item Docker might require sudo access.
  \item Email updates can be obtained by providing a .env file in the root folder; this requires the generation of a Google app password.
\begin{verbatim}
EMAIL_ADDRESS=email@gmail.com
EMAIL_PASSWORD=16stringpassword
\end{verbatim}
\end{enumerate}

\end{document}